# Planar Sensor for RF Characterization of magnetic samples

Nilesh K Tiwari, *Student Member, IEEE*, A K Jha, *Member, IEEE*, S P Singh, *Student Member, IEEE*, M Jaleel Akhtar, *Senior Member, IEEE*

*Abstract*—A magnetic measurement of the bar shaped test specimen placed inside the planar sensor is presented. A magnetic material characterization approach using planar cavity is proposed in this work. The proposed planar sensor relaxes the main limitations of conventional approach by using the proper feeding section. The proposed sensor is numerically verified using the full wave EM simulator for the magnetic property estimation. It is found that the developed sensor is able to characterize the test specimen with improved accuracy than that of conventional approach.

## I. Introduction

THE perturbation formulation especially the material perturbation approach actually provides the equations relating the sensor parameters with the constitutive electromagnetic parameter of the test specimen [1]-[5]. The designed resonant sensor in literature basically operates quite close to its dominant mode and not being much tested for magnetic samples. It is mainly due to the aforementioned reasons that in this work the design of planar magnetic sensor is presented similar to [6]-[7].

## II. Analytical Formulation

The full wave electromagnetic solver CST-MWS is employed to perform the numerical analysis of the proposed sensor design [6]. The test specimen basically helps to perturb the horizontal magnetic field of planar sensor as the electric field remains absent at the center of SIW cavity.

However, form the absolute magnetic field plot it can be visualize that the maximum of magnetic field appears at the center of SIW cavity. From this figure one can easily be observed that the horizontal placement of sample along the width of test specimen facilitates its interaction with the magnetic field component only.

The magnetic property of the test specimen placed horizontally at the center of planar cavity can be related with the resonant frequency ad quality factor of the cavity using the perturbation formulation. The simple material perturbation relation corresponding to the test specimen with $\mu_r = \mu_r' - j\mu_r''$ loaded on the planar cavity designed on the substrate with complex permeability can be given as [3], [6]:

$$\frac{f - f_0}{f} = -\frac{\frac{1}{2}\iiint_{V_{sample}}\left(\frac{\mu_r}{ } - 1\right)\vec{H}_0^* \cdot \vec{H}\, dv}{\iiint_{V_C}\left|\vec{H}_0\right|^2 dv} \quad (1)$$

Now from the above equation corresponding to even $TE_{102n}$ modes it can be noticed that in the present case only the $H_z$ component of magnetic field actually interact with the test specimen. These values of magnetic field component can now be used in (1) to recalculate the shift in complex resonant frequency over the finite sample volume. The relation given in (1) is now modified for the $4^{th}$ even mode n=2 as (2) which can be used to test the magnetic sample.

$$\frac{f - f_0}{f} = -\left(\frac{\mu_r}{2\mu_{rS}} - \frac{1}{2}\right) 4a^2 \left(1 - \text{sinc}(k_z a_1)\right)\left(4\pi a^2 + \lambda_g^2\right)^{-1} \left(1 - \text{sinc}(k_x l_1)\right) \quad (2)$$

## III. NUMERICAL MAGNETIC PROPERTY CALCULATION

The relationship (2) is now used to calculate the permeability of test specimen at particular frequency point using the numerically generated S-parameters. The CST-MWS is used for the generation of scattering parameters corresponding to different samples defined in the CST material library. The numerically generated scattering parameters are then used to record the resonant frequency and quality factor. The typical plots of scattering coefficient at 4th even modes are shown in Fig.1.

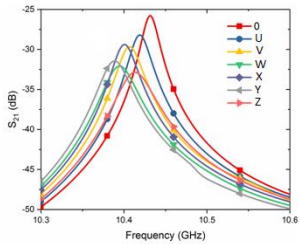

(a)

Fig. 1 plot of scattering parameters for samples U-Z for $TE_{104}$.

The estimated value of magnetic property of samples is then tabulated in Table I. The Table I include the information about estimated and actual value of real permeability and magnetic loss tangent of test samples. It is found that the calculated value of complex permeability using the proposed approach is relatively closer to their actual value

Table I: COMPARISON BETWEEN THE COMPLEX PERMEABILITIES OBTAINED WITH THE CONVENTIONAL APPROACH (subscript C) AND THE MODIFIED FORMULA FOR n=2

| Test Samples | Actual $\mu'_r$ | $\mu'_{rP}$ | Actual $\tan\delta_m$ | $\tan\delta_{mP}$ |
|---|---|---|---|---|
| U | 1.2 | 1.206 | 0.040 | 0.0397 |
| V | 1.4 | 1.398 | 0.060 | 0.0602 |
| W | 1.6 | 1.593 | 0.100 | 0.1034 |
| X | 1.5 | 1.495 | 0.050 | 0.0499 |
| Y | 1.7 | 1.691 | 0.008 | 0.0814 |
| Z | 1.3 | 1.302 | 0.150 | 0.1584 |

* Here test samples U –Z represents the materials considered for numerical simulation.

## IV. CONCLUSION

The planar sensor based resonant approach for magnetic characterization of test samples has been derived. The proposed sensor proves to be quite reasonable than that of its conventional counterpart. The proposed sensor is verified numerically using the obtained S-parameters corresponding to sample loading.

REFERENCES

[1] K. T. Mathew, "Perturbation theory," in *Encyclopedia of RF andMicrowave Engineering*, vol. 4. New York, NY, USA: WileyInterscience, 2005, pp. 3725–3735.
[2] R. A. Waldron, "Perturbation theory of resonant cavities," *Proc. Inst.Elect. Eng.*, vol. 107, pp. 272–274, Sep. 1960 .
[3] R. F. Harrington, *Time-Harmonic Electromagnetic Fields*. New York,NY, USA: Wiley-Interscience, 2001, pp. 317–349.
[4] U. Raveendranath and K. T. Mathew, "New cavity perturbation techniquefor measuring complex permeability of ferrite materials," *Microw. Opt.Technol. Lett.*, vol. 18, no. 4, pp. 241–243, Jul. 1998.
[5] L. F. Chen, C. K. Ong, C. P. Neo, V. V. Varadan, and V. K. Varadan,*Microwave Electronics: Measurement and Materials Characterization*.London, U.K.: Wiley, 2004.
[6] A. K. Jha and M. J. Akhtar, "An improved rectangular cavity approachfor measurement of complex permeability of materials," *IEEE Trans.Instrum. Meas.*, vol. 64, no. 4, pp. 995–1003, Apr. 2015.